\documentclass[a4paper,12pt]{article}
\pdfoutput=1
\usepackage{graphicx,rotating,hyperref,slashed,amsmath,xcolor,amssymb,amsfonts,colortbl,cite} 
\makeatletter
\hypersetup{colorlinks,bookmarksopen,bookmarksnumbered,
linkcolor=blus,pdfstartview=FitH,urlcolor=rossos,citecolor=verde}
\allowdisplaybreaks	 
\numberwithin{equation}{section}
\hypersetup{%
    ,urlcolor=black
    ,citecolor=black
    ,linkcolor=black
    }
\usepackage{mciteplus}

\AtBeginDocument{% optionally set colors to your liking
  \hypersetup{
    urlcolor=blue,
    citecolor=blue,
    linkcolor=black,
  }%
}

\def\beq{\begin{equation}}
\def\eeq{\end{equation}}

\newcommand\ees{\end{eqnarray}}
\newcommand\bees{\begin{eqnarray}}

\def\bea{\begin{eqnarray}}
\def\eea{\end{eqnarray}}

\def\d{{\rm d}}

\def\d{{\rm d}}

\def\0{{\boldsymbol 0}}

\def\lsim{\mathrel{\rlap{\lower3pt\hbox{\hskip0pt$\sim$}}
   \raise1pt\hbox{$<$}}}         %less than or approx. symbol
\def\gsim{\mathrel{\rlap{\lower4pt\hbox{\hskip1pt$\sim$}}
   \raise1pt\hbox{$>$}}}         %greater than or approx. symbol

 \newcommand{\sfootnote}[1]{}
\definecolor{bluc}{cmyk}{1,1,0,0.1}
\definecolor{rossoCP3}{cmyk}{0,.88,.77,.40}
\definecolor{rosso}{cmyk}{0,1,1,0.4}
\definecolor{rossos}{cmyk}{0,1,1,0.55}
\definecolor{rossoc}{cmyk}{0,1,1,0.2}
\definecolor{verdes}{cmyk}{0.92,0,0.59,0.4}

\hypersetup{colorlinks, bookmarksopen, bookmarksnumbered,
citecolor=verdes, linkcolor=bluc, pdfstartview=FitH, urlcolor=rossos}

\usepackage{multicol}
\usepackage{color}
\definecolor{rosso}{cmyk}{0,1,1,0.4}
\definecolor{rossos}{cmyk}{0,1,1,0.55}
\definecolor{rossoc}{cmyk}{0,1,1,0.2}
\definecolor{blu}{cmyk}{1,1,0,0.3}
\definecolor{blus}{cmyk}{1,1,0,0.6}
\definecolor{bluc}{cmyk}{1,1,0,0.1}
\definecolor{verde}{cmyk}{0.92,0,0.59,0.25}
\definecolor{verdec}{cmyk}{0.92,0,0.59,0.15}
\definecolor{verdes}{cmyk}{0.92,0,0.59,0.4}

\skip\@mpfootins = \skip\footins
\renewcommand\&{&}

\def\circa#1{\,\raise.3ex\hbox{$#1$\kern-.75em\lower1ex\hbox{$\sim$}}\,}

\newcommand{\be}{\begin{equation}}
\newcommand{\ee}{\end{equation}}
\newfam\rsfsfam
\def\mathscr#1{{\fam\rsfsfam\relax#1}}

\def\circa#1{\,\raise.3ex\hbox{$#1$\kern-.75em\lower1ex\hbox{$\sim$}}\,}
\makeatletter

\def\hhref#1{\href{http://arxiv.org/abs/#1}{arXiv:#1}} % in bibliography

\newcommand{\doi}[1]{\href{http://dx.doi.org/#1}{[doi]}}

\def\hhref#1{\href{http://arxiv.org/abs/#1}{arXiv:#1}} 
 
\def\art{\@ifnextchar[{\eart}{\oart}}
\def\eart[#1]#2#3#4#5#6{{\rm #2}, {\em #3 \bf #4} {\rm (#6) #5} ({\em #1})}

\def\article{\@ifnextchar[{\earticle}{\oarticle}}
\def\oarticle#1#2#3#4#5#6{{\rm #1}, {\em ``#6''}, {\rm #2 #3 (#5) #4}}
\def\earticle[#1]#2#3#4#5#6#7{{\rm #2}, {\em ``#7''}, {\rm #3 #4 (#6) #5}  [\hhref{#1}]}
\def\hepart[#1]#2{{\rm #2, \em#1}}
\def\heparticle[#1]#2#3{#2, {\em ``#3''} [\hhref{#1}]}

\newcounter{alphaequation}[equation]
\def\thealphaequation{\theequation\hbox to
0.6em{\hfil\alph{alphaequation}\hfil}}
\def\eqnsystem#1{
\def\@eqnnum{{\rm (\thealphaequation)}}
\def\@@eqncr{\let\@tempa\relax \ifcase\@eqcnt \def\@tempa{& & &} \or
  \def\@tempa{& &}\or \def\@tempa{&}\fi\@tempa
  \if@eqnsw\@eqnnum\refstepcounter{alphaequation}\fi
\global\@eqnswtrue\global\@eqcnt=0\cr}
\refstepcounter{equation} \let\@currentlabel\theequation \def\@tempb{#1}
\ifx\@tempb\empty\else\label{#1}\fi
\refstepcounter{alphaequation}
\let\@currentlabel\thealphaequation
\global\@eqnswtrue\global\@eqcnt=0 \tabskip\@centering\let\\=\@eqncr
$$\halign to \displaywidth\bgroup \@eqnsel\hskip\@centering
$\displaystyle\tabskip\z@{##}$&\global\@eqcnt\@ne
\hskip2\arraycolsep\hfil${##}$\hfil& \global\@eqcnt\tw@\hskip2\arraycolsep
$\displaystyle\tabskip\z@{##}$\hfil
\tabskip\@centering&\llap{##}\tabskip\z@\cr}
\def\endeqnsystem{\@@eqncr\egroup$$\global\@ignoretrue} \makeatother

\definecolor{fiorentina}{rgb}{.5,0,.5}

\begin{document}
%\preprint{ET-0465A-21}

%ET-0465A-21
\setcounter{page}{1} \baselineskip=15.5pt \thispagestyle{empty}

\vspace{0.8cm}
\begin{center}

{\fontsize{19}{28}\selectfont  \sffamily \bfseries {A large $|\eta|$ approach to single field inflation}}

\end{center}

\vspace{0.2cm}

\begin{center}
{\fontsize{13}{30}\selectfont  Gianmassimo Tasinato$^{1,2}$ } 
\end{center}

\begin{center}

\vskip 8pt
\textsl{$^{1}$ Dipartimento di Fisica e Astronomia, Universit\`a di Bologna,  Italia}
\\
\textsl{$^{2}$ Physics Department, Swansea University, SA28PP, United Kingdom}\\
\textsl{\texttt{email}: g.tasinato2208 at gmail.com }\\
\vskip 7pt

\end{center}

\smallskip
\begin{abstract}
\noindent
Single field models of inflation capable to produce primordial black holes usually require
a significant departure from the standard, perturbative slow-roll regime. In fact, in many of these scenarios, the size of the slow-roll parameter  $|\eta|$  becomes  larger than one
during a  short phase  of  inflationary evolution. In order 
to develop an analytical control on these systems, 
  we  explore the  limit of $|\eta|$  large, and  promote  $1/|\eta|$ to a small quantity
  to be   used  for  
perturbative expansions. 
Formulas simplify, and we obtain
analytic expressions for  the two and three point functions  of curvature fluctuations, which 
 share  some of  the  features  found in  realistic inflationary models generating  primordial black holes.  
We study one-loop corrections in this  framework: we discuss criteria for adsorbing
ultraviolet divergences into the available parameters, leaving log-enhanced
infrared contributions of controllable size.
\end{abstract}

\section{Introduction and Conclusions}
\label{sec_intro}

Identifying the nature of dark matter is 
one of the most challenging open problems in cosmology \cite{Bertone:2016nfn}.
A fascinating possibility is that dark matter is made of primordial
black holes (PBH) \cite{Hawking:1971ei,Carr:1974nx,Ivanov:1994pa,Garcia-Bellido:1996mdl}, forming from the collapse of high density fluctuations
produced during cosmic inflation: see e.g. \cite{Khlopov:2008qy,Garcia-Bellido:2017fdg,Sasaki:2018dmp,Carr:2020xqk,Green:2020jor,Escriva:2022duf,Ozsoy:2023ryl} for reviews. In order  for producing PBH, the size of the inflationary curvature fluctuation spectrum
needs to increase by around seven orders of magnitude, from large to small scales. This
condition is not possible to achieve within a controlled slow-roll expansion in  single-field inflation \cite{Motohashi:2017kbs}: a departure from the standard slow-roll conditions is needed. In several
single-field realizations of PBH scenarios, the size $|\eta|$
of the second slow-roll parameter becomes larger than
one during a brief phase of non-slow-roll evolution (from now on, NSR). Such brief NSR era should last few e-folds $\Delta N_{\rm NSR}$ of expansion. Examples are ultra-slow-roll models \cite{Kinney:2005vj,Martin:2012pe,Dimopoulos:2017ged}, where $\eta=-6$, and constant roll models \cite{Motohashi:2014ppa,Inoue:2001zt,Tzirakis:2007bf}, where $|\eta|$ can be larger or smaller than $6$, depending on the properties of the inflationary potential. In  these cases, the  evolution of fluctuations
 challenges   analytical investigations, since the slow-roll expansion breaks down. Wands duality \cite{Wands:1998yp} can be of help in the ultra-slow-roll case, but still care is needed in connecting slow-roll to NSR eras. Oftentimes, a numerical analysis is needed. 

\smallskip
 
In this work, we  consider large values for the slow-roll quantity 
 $|\eta|$, and use the inverse
$1/|\eta|$ as expansion parameter. A large value of $|\eta|$ is not inconceivable to obtain
at the price of tunings, for example in constant roll systems. Here we are not interested
in model building, but in investigating the consequences of a large $|\eta|$ limit
for the dynamics of fluctuations. When working at leading order in $1/|\eta|$ formulas simplify, and we obtain analytic expressions for the two and three point functions of curvature fluctuations. These analytic results can be useful to get  insights
on the properties of curvature fluctuations in PBH scenarios, as well as  understanding 
  the physical
consequences of a rapid growth of the curvature spectrum from large to small scales. 

\smallskip

This idealized, large-$|\eta|$ limit has some intriguing analogy with the large-$N$ limit 
of $SU(N)$ QCD, a model introduced by  't Hooft \cite{tHooft:1973alw} in a
 particle physics context. $N$ being the number of colors,  the field-theory analysis can be carried on using a perturbative $1/N$ expansion, and simplifies in a large-$N$ limit. Real world QCD has $N=3$ colors only, yet the results of an $1/N$ expansion  catch various important
properties of standard QCD: we refer the reader to chapter 8 of \cite{Coleman:1985rnk} for a pedagogical
survey. Calling $g$ the QCD coupling constant, and $N$ the number of colors,   't Hooft finds   convenient to take the simultaneous limits $g\to0$, $N\to \infty$, and $g N^2$ fixed \cite{tHooft:1973alw}. Analogously, in PBH forming scenarios, it is convenient to consider the limit of vanishing e-folds of NSR expansion,  $
\Delta N_{\rm NSR}\to0$, and
at the same time taking $|\eta|\to\infty$, keeping
 fixed the product $\left( \Delta N_{\rm NSR} |\eta| \right)$. As we will learn, 
this product is associated with  the growth of the spectrum from large to small scales. Keeping  
$\left( \Delta N_{\rm NSR} |\eta| \right)$ fixed, and expanding in $1/|\eta|$, the formulas for
the curvature fluctuation $n$-point functions become  easier to deal with.

Having analytical control on a perturbative expansion in $1/|\eta|$
allows us to address the issue of loop corrections,
a topic that recently raised much attention after the important papers  \cite{Kristiano:2022maq,Kristiano:2023scm}
appeared. As pointed out in \cite{Kristiano:2022maq},  the same mechanisms
that causes the curvature spectrum growth, as needed for producing PBH,  also
amplify the effects of loop corrections to the curvature power spectrum. Their size can become  so large to invalidate a perturbative loop expansion. Many solutions
and new perspectives have recently pointed out \cite{Riotto:2023hoz,Choudhury:2023vuj,Choudhury:2023jlt,Riotto:2023gpm,Choudhury:2023rks,Firouzjahi:2023aum,Motohashi:2023syh,Choudhury:2023hvf,Firouzjahi:2023ahg,Tomberg:2023kli,Firouzjahi:2023btw,Franciolini:2023lgy}.  
In the framework of a large $|\eta|$ expansion,
we show that
 loop corrections
can be placed under control, at least at the  large scales that
can affect CMB physics.   We  regularize loop integrals by means of ultraviolet and infrared cut-offs, and  analytically compute
the effects of loops in a large $|\eta|$ regime. The resulting ultraviolet divergences can be adsorbed into physically measurable quantities corresponding to the amplitude and the large-scale tilt of the spectrum. We 
are left
 with log-enhanced
infrared contributions, whose size  is small at large scales.

\smallskip

We hope that the tool of 
an $1/|\eta|$ expansion, although idealized, can lead to 
analytical insights allowing to further  investigate  properties of the dynamics of curvature fluctuations
in PBH scenarios. It will be interesting to further apply this method to related topics, as the behaviour of higher order $n$-point functions, and their corresponding loop corrections in a large $|\eta|$ limit.  Having analytic expressions for the primordial correlators can also be useful for investigating the actual process of PBH formation in the post-inflationary universe, as well as the generation of second-order gravitational waves from enhanced curvature spectra: see respectively e.g. \cite{Escriva:2021aeh} and \cite{Domenech:2021ztg} for reviews. We leave these topics to future investigations. 

\newpage
%%%%%%
\section{System under consideration}
\label{sec_system}

We consider single field models of inflation with canonical kinetic terms. Around a conformally
flat cosmological  metric,
$d s^2\,=\,a^{2}(\tau) \left(-\d \tau^2+d \vec x^2\right)$,
 the quadratic action for the
curvature perturbation %$\zeta_k(\tau)$
in  Fourier space reads (we set the Planck mass to unity)
\bea
\label{quadac}
S_{\rm quad}\,=\,\frac12 \int d \tau\,d^3  k\,{z^2(\tau)}\left[\zeta_k'^2(\tau)+
k^2 \zeta_k^2(\tau)\right]\,,
\eea
where the pump field $z(\tau)$ is given by
\be
z(\tau)\,=\,a(\tau) \,\sqrt{2\epsilon (\tau)}\,.
\ee
The definitions of  Hubble and slow-roll parameters are
\bea
H(\tau)\,=\,\frac{a'(\tau)}{a^2(\tau)}\hskip0.7cm;\hskip0.7cm \epsilon(\tau)\,=\,-\frac{H'(\tau)}{a(\tau)\,H^2(\tau)}
\hskip0.7cm;\hskip0.7cm \eta(\tau)\,=\,\frac{\epsilon'(\tau)}{a(\tau)\,H(\tau)\,\epsilon(\tau)}
\,.
\label{defsr1}
\eea
We assume that the first slow-roll parameter $\epsilon(\tau)$ remains  small during the entire duration of inflation, 
which takes place for negative conformal time $\tau\le \tau_0\,=\,0$. We also
assume that 
the second parameter
$\eta(\tau)$
 remains  small for negative values of $\tau$,  
a
part from a brief time interval  $\tau_1\le \tau \le \tau_2$ during which 
$\eta$ is negative and 
its size $|\eta|$ becomes larger
than one. (See the
brief discussion in Section \ref{sec_intro}.) During this short phase, which we call non-slow-roll
(NSR) period,  
we  can not  make a perturbative slow-roll expansion
in  $|\eta|$: other methods are needed to tackle the evolution
of fluctuations.
 In this work, we explore the possibility   to consider the inverse $1/|\eta|$
 as a convenient expansion parameter for pursuing analytical
 considerations. But before discussing the role of the $|\eta|$ parameter, we first
 examine a quantity  related with the duration of NSR phase.
   We build a dimensionless positive parameter $\Delta \tau$, as
\be \label{defDT}
\Delta \tau\,=\,-\frac{\tau_2-\tau_1}{\tau_1}\,,
\ee
and we require that $\Delta \tau\ll1$. This condition implies that the duration of the NSR
phase is  short with respect to the typical time-scales
 one encounters in treating the system, as e.g.
 $|\tau_1|$ which controls
 the onset of the NSR phase. A short duration of non-slow-roll phase is demanded by the requirement to avoid
 excessive stochastic effects \cite{Pattison:2017mbe,Ezquiaga:2019ftu,Figueroa:2021zah}. 
 Since we assume that the slow-roll parameter $\epsilon(\tau)$
remains always small during inflation, we  consider 
for simplicity 
the limit of pure de Sitter expansion,
with $a(\tau)\,=\,-1/(H_0 \tau)$  and $H_0$
constant during inflation. If the interval $\Delta \tau$ of eq \eqref{defDT} is small, this parameter
has a  physical interpretation in terms of a
% corresponds to
 (small) number $\Delta N_{\rm NSR}$ of e-folds of NSR evolution:
\bea
\Delta N_{\rm NSR}&=&\ln{\left(
\frac{a(\tau_2)}{a(\tau_1)}
\right)
}\,=\,\ln\left({\frac{\tau_1}{\tau_2}}\right)\,=\,\ln\left( \frac{1}{1-\Delta \tau}\right)\,\simeq\,\Delta \tau\,,
\eea
where in the next-to-last equality we used the definition \eqref{defDT}, and in the last equality we expanded for small $\Delta \tau$. 

In the regime of $\Delta \tau \ll1$ we can use the results of \cite{Tasinato:2020vdk} (reviewed in the
technical Appendix \ref{AppA}): we write
the solution for the mode function of the curvature perturbation $\zeta_{\kappa}(\tau)$ in Fourier space
during different epochs in the inflationary evolution.
We define the pivot scale 
\be
\label{defpiva}
k_\star\,=\,1/|\tau_1|\,,
\ee
 corresponding to modes leaving the horizon
at the onset of the NSR era.
We
  express our formulas in terms
of  dimensionless momentum scales, as follows:
\be \label{dimkd}
\kappa\equiv -k \tau_1\,=\,{k}/{k_\star}\,.
\ee
Our expressions simplify with  this notation, as we can easily
 identify  modes  with  $\kappa\sim1$ which cross the horizon at epochs corresponding to the
 NSR phase. For this reason, we adopt from now on the dimensionless definition \eqref{dimkd} when treating momenta. 
 
The mode function $\zeta_\kappa (\tau)$ acquires its usual profile matching  the
Bunch-Davies vacuum at  short distances:  
\be
\label{modeSR}
\zeta_\kappa (\tau)\,=\,-i\,\frac{H_0\,(-\tau_1)^{3/2}}{\sqrt{4 \epsilon_1}\,\kappa^{3/2}}\left(1-\frac{i \tau}{\tau_1}\right)\,e^{i \frac{\kappa\tau}{\tau_1}}
\hskip0.8cm;\hskip0.8cm \tau\le \tau_1
\ee
for conformal times $\tau\le\tau_1$, since at early times the modes  do not yet experience
 the NSR evolution.
In the previous equation, $H_0$ is the constant Hubble parameter during inflation, and
\be
\epsilon_1\,=\,\epsilon(\tau_1)
\ee
 is the value of the first slow-roll parameter at  $\tau=\tau_1$.   

  For later times $\tau_2\le\tau\le\tau_0$ during inflation, instead, the profile of the mode function is modified
  by the effects of the NSR era.  See Appendix \ref{AppA}, where
  we  include the behaviour of the mode function in the interval $\tau_1\le\tau\le \tau_2$ that we do not need in the main text. (Sufficient to say
  that the mode functions, with their first derivatives, are continuous at the transition
  between slow-roll and non-slow-roll eras.) We find  %{\bf \color{red} substitute 8 with 4}
\be
\label{modeNSR}
\zeta_k (\tau)\,=\,-i\,\frac{H_0\,(-\tau_1)^{3/2}}{\sqrt{4 \epsilon_1}\,\kappa^{3/2}}
\left[{\cal C}_1(\kappa)\,
\left(1-\frac{i \tau}{\tau_1}\right)\,e^{i \frac{\kappa\tau}{\tau_1}}
+{\cal C}_2(\kappa)\,
\left(1+\frac{i \tau}{\tau_1}\right)\,e^{-i \frac{\kappa\tau}{\tau_1}}
\right]
\hskip0.3cm;\hskip0.3cm \tau_2\le \tau\le\tau_0
\ee
with (recall the definition of $\Delta \tau$ in eq \eqref{defDT})
%{\bf \color{red} check sign of eta}
\bea
{\cal 
C}_1(\kappa)&=&1-\frac{\eta}{8 (1-\Delta \tau)^2\,\kappa^2} \left[ 1-e^{2 i \kappa \Delta \tau}
-2 \kappa \Delta \tau \left(i-2 \kappa (1-\Delta \tau) \right)
\right]\,,
\label{defC1}
\\
{\cal C}_2(\kappa)&=&
\frac{\eta\,e^{2 i (1-\Delta \tau) \kappa}}{8 (1-\Delta \tau)^2\,\kappa^2} \left[ 1
- 2 i \kappa-
e^{2 i \kappa \Delta \tau}
 \left(1-2i \kappa (1-\Delta \tau) \right)
\right]\,,
\label{defC2}
\eea
and 
\be
\label{etadef}
\eta\,=\,\lim_{\tau\to\tau_1^+}\eta(\tau)
\,.
\ee
From now on,  the quantity  $\eta$ refers to the definition \eqref{etadef},
i.e. the value of the time-dependent $\eta(\tau)$ evaluated at the beginning of the NSR era. Notice that in the limit of negligible $\eta\to0$, the two mode
functions \eqref{modeSR} and \eqref{modeNSR} coincide.  Instead, if $|\eta|$
is large in size, the scale dependence of the mode functions \eqref{modeSR}  and  \eqref{modeNSR}
differs considerably. This leads to the opportunity of increasing the size of the curvature
spectrum at small scales, as required by primordial black hole production.
In what comes next, 
we examine this possibility.

\section{The two-point function of curvature fluctuations}
\label{sec_power}

\begin{figure}
\begin{center}
    \includegraphics[width = 0.49 \textwidth]{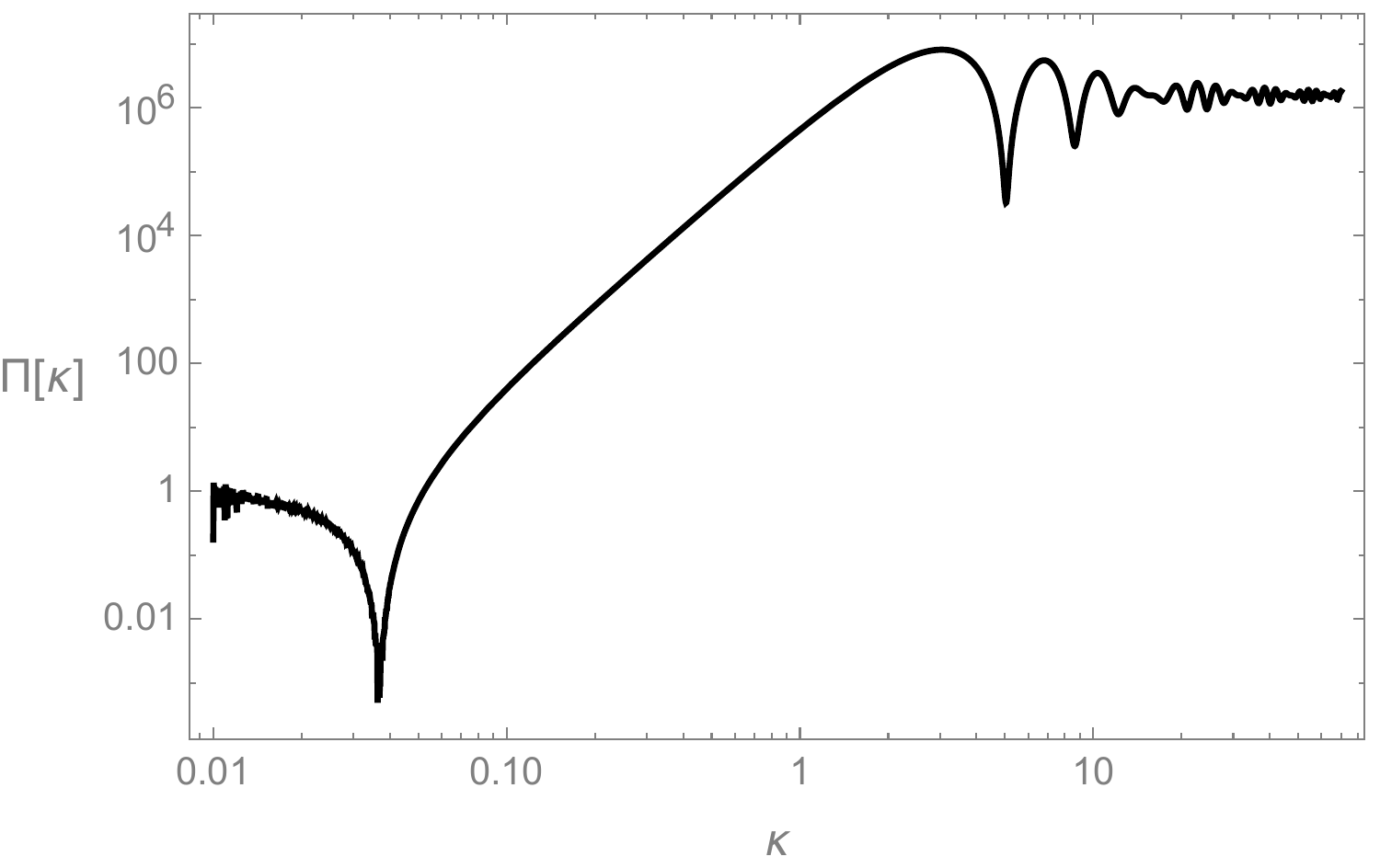}
  \includegraphics[width = 0.49 \textwidth]{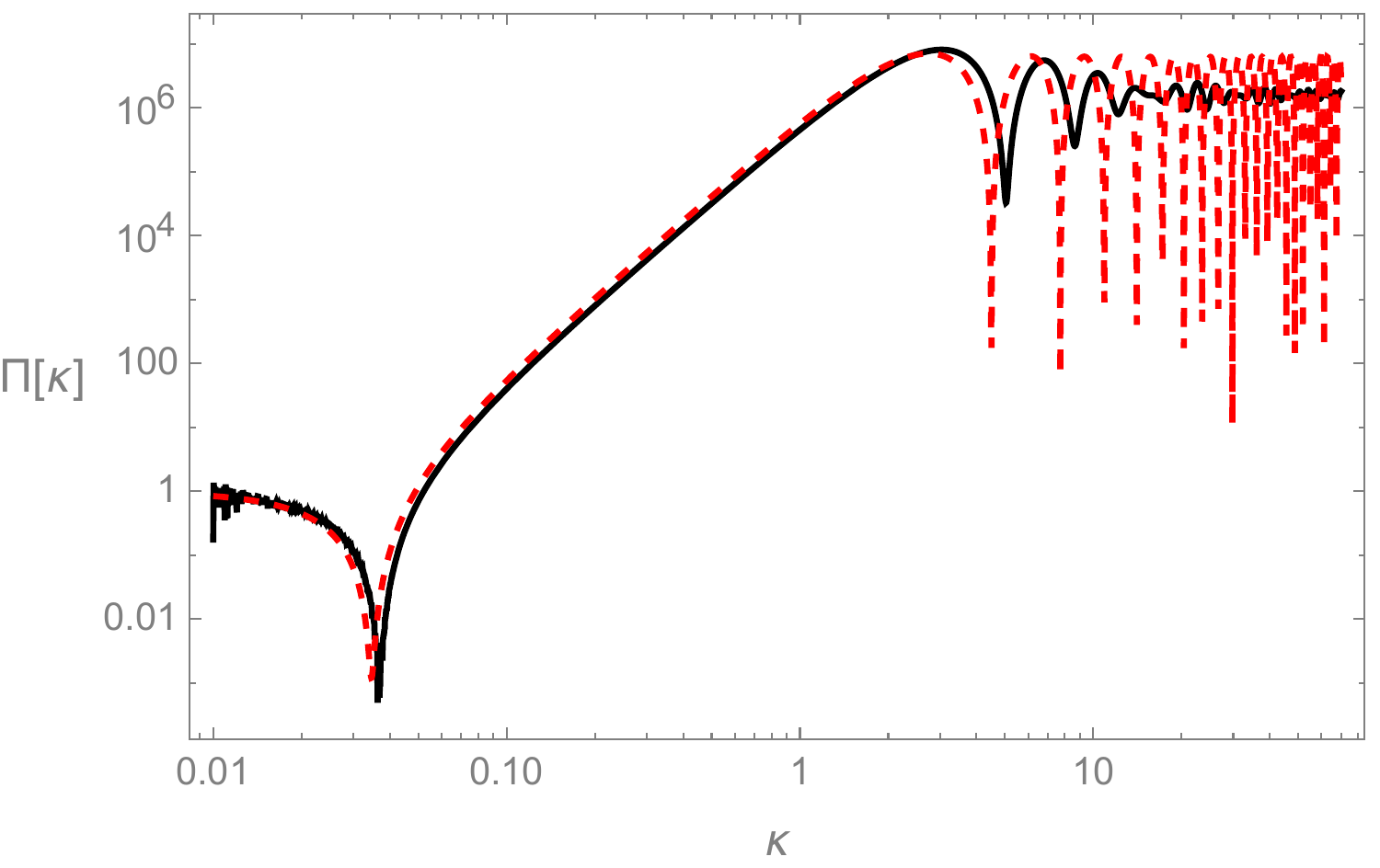}
    \end{center}
 \caption{ \small {\bf Left panel:} Plot of the dimensionless power spectrum ${ \Pi}(\kappa)$
 as defined in eq \eqref{finexpi}: we choose the values  
  $|\eta|=10^4$ and $\Delta \tau=0.2$ for the free parameters. {\bf Right panel}: the black line
  is the same as left panel. The dashed red line represents the spectrum $\hat{\Pi}(\kappa)$ of eq \eqref{simPS}, choosing the value 
$\Pi_0=1250$ for the single free parameter. See  the discussion
after eq \eqref{defj1}. Notice that the maximal values of the spectrum occur around the onset 
of non-slow-roll phase, for $\kappa \sim {\cal O}(1)$.} 
  \label{fig_confinfA}
\end{figure}

In this section we show how a suitably defined large-$|\eta|$ limit  allows us to analytically capture
the  scale dependence of the spectrum of curvature fluctuations. Starting from the mode functions obtained in the previous Section, we  quantize the system starting from
the 
 quadratic action \eqref{quadac}
for curvature fluctuations. See e.g. \cite{Liddle:2000cg} for 
a textbook discussion. We can easily compute the two-point function 
$\langle \zeta_\kappa(\tau_0)\, \zeta_\kappa^*(\tau_0)\rangle$
of curvature
perturbations evaluated at the end of inflation, $\tau=\tau_0=0$, and the corresponding
power spectrum (recall our definition \eqref{dimkd} of dimensionless scale $\kappa$)
\be \label{def_spec}
{\cal P}_{\kappa}\,\equiv\,\frac{\kappa^3}{2 \pi^2(-\tau_1)^3}\,\langle \zeta_\kappa(\tau_0)\, \zeta_\kappa^*(\tau_0)\rangle'\,,
\ee
where a prime indicates the two-point function omitting the momentum-conserving delta functions.
At  very large scales, $\kappa\to0$, one finds the usual expression
\be\label{lsspec}
{\cal P}_{0}\,=\,\lim_{\kappa\to0 }{\cal P}_{\kappa}\,=\,\frac{H_0^2}{8\pi^2 \epsilon_1}
\,,
\ee
with the scale of ${\cal P}_{0}$ of order $10^{-9}$ to match CMB normalization. 
Since large scale modes leave the horizon much earlier than the NSR era,
they are unaffected by it. 
 It is convenient to compute
the dimensionless ratio $\Pi(\kappa)$ (see \cite{Tasinato:2020vdk})
between the power spectrum \eqref{def_spec} evaluated at scale $\kappa$,  versus the large-scale 
spectrum ${\cal P}_0$ in eq \eqref{lsspec}.
 We find
\be
\label{finexpi}
\Pi(\kappa)\,\equiv\,
\frac{{\cal P}_\kappa }{\lim_{\kappa\to0} {\cal P}_\kappa}\,=\,|{\cal C}_1(\kappa)+{\cal C}_2(\kappa)|^2
\,,
\ee
with the scale-dependent ${\cal C}_{1,2}(\kappa)$ given in eqs \eqref{defC1}, \eqref{defC2}.  
Such ratio  can be considered as a dimensionless power spectrum
evaluated at the end of inflation, which
singles out the overall amplitude ${\cal P}_0$ at large scales, and encampsulates the 
rich scale dependence of the spectrum evolving from large to small scales. We plot  $\Pi(\kappa)$ in Figure \ref{fig_confinfA}, left panel, for a representative
choice of parameters capable to enhance the spectrum at small scales.   Physically, the
scale dependence of the spectrum is due to the brief 
  NSR phase of inflationary evolution.  
The NSR era is able to excite the would-be
decaying mode at superhorizon scales, which starts  to actively participate to the dynamics
of curvature fluctuations.  
 See e.g.
\cite{Ozsoy:2023ryl} for a recent review.
Notice that the spectrum  has  a pronounced dip at intermediate scales, 
due to  a disruptive interference between the growing
and decaying modes of  the curvature fluctuation at super-horizon scales. The dip is followed
by a steady growth (with slope $\kappa^4$ as first shown in \cite{Byrnes:2018txb}) until it reaches a maximal amplitude.  
See also \cite{Ozsoy:2019lyy} for a detailed analysis of the shape of the curvature power spectrum in PBH 
forming scenarios. We point out that -- while in this work we evaluate all quantities at the end of inflation, when the inflationary
dynamics ceases to affect the evolution of curvature fluctuations -- the large $|\eta|$ approach can be also applied to
compute correlators at any time during the inflationary era. 

\bigskip

It is particularly interesting to evaluate the value of $\Pi(\kappa)$ at very small scales, $\kappa\to\infty$, which informs us on the total amount   of the growth of  the
spectrum. See  Figure \ref{fig_confinfA}, left panel. Plugging into \eqref{finexpi}
the expressions for ${\cal C}_{1,2}$ of eqs \eqref{defC1}, \eqref{defC2} and taking the small-scale limit, we find
\bea
\lim_{\kappa\to\infty}\Pi(\kappa)
&=&\left(\frac{1+\left(|\eta|/2-1\right)\,\Delta \tau}{1-\Delta \tau}\right)^2
\,,
\nonumber
\\
&\equiv&(1+\Pi_0)^2\,,
\label{defPI}
\eea
where in the second line we introduce a constant parameter $\Pi_0$ controlling the 
enhancement of the spectrum from large to small scales ($\Pi_0=0$ means no enhancement).
 We would like a large enhancement of the spectrum at small scales for producing PBH. 
Since
 we are in a regime of small $\Delta \tau$, as discussed in Section \ref{sec_system},  we  need to consider large values
for the parameter $|\eta|$  during the NSR period (we make the hypothesis that $\eta$ is negative, hence the absolute value). In fact,
in the limit of $|\eta|$ large and  $\Delta \tau$ small,  expression \eqref{defPI} simplifies to
%If large enhancement, we have expression
\be
\label{fixedPI}
\Pi_0\,\simeq\,\frac{|\eta| \Delta \tau}{2}\,.
\ee
The combination \eqref{fixedPI}, as well as the considerations above, motivates us to take  the simultaneous limits:
\be
\label{deflimit}
|\eta| \gg 1\hskip0.7cm;\hskip0.7cm \Delta \tau\ll1  \hskip0.7cm;\hskip0.7cm {\rm keep\,\,\,\,} \Pi_0\,\,\,\,{\rm fixed\,.}
\ee
 This is reminiscent of the  't Hooft limit one encounters in particle physics \cite{tHooft:1973alw},
as explained in Section \ref{sec_intro}. In fact, combining
$|\eta|$ and $ \Delta \tau$ into the fixed quantity  $ \Pi_0$  allows us to consistently 
 perform
  expansions in the small parameter
$1/|\eta|$,  maintaining at the same time control on the effects of the NSR through
the quantity $\Pi_0$.
 In most PBH scenarios we aim to a total enhancement of the order $10^6-10^7$
 in eq \eqref{defPI}. Then the quantity 
 $\Pi_0$ results  by itself large, of order $10^3-10^4$. 

\bigskip

Adopting
 the limits
 of eq \eqref{deflimit}, the expression for the  ratio \eqref{finexpi} simplifies. We substitute $\Delta \tau\,=\,2 \Pi_0/ |\eta| $ in eq \eqref{finexpi}, and expand  for large values of $|\eta|$
 keeping $\Pi_0$ fixed.  At leading
 order in this expansion, we obtain %the concise analytic expression
\be
\label{simPS}
\hat { \Pi}(\kappa)\,=\,1-4 \kappa\,\Pi_0\,\cos{\kappa}\,j_1(\kappa)
+4 \kappa^2 \,\Pi_0^2\,j^2_1(\kappa)\,,
\ee
where a hat indicates  that we only include the leading order in
an expansion in $1/|\eta|$, following the conditions of eq \eqref{deflimit}.  The spherical Bessel function $j_1(\kappa)$ is given by
\be
\label{defj1}
 j_1(\kappa)\,=
\, \frac{\sin \kappa}{\kappa^2}-\frac{\cos \kappa}{\kappa}\hskip0.7cm;\hskip0.7cm j_1(\kappa\ll1)\,=\,\frac{\kappa}{3}-\frac{\kappa^3}{30}+{{\cal O}(\kappa^5)}\,.
\ee
We represent formula 
 \eqref{simPS} in Figure \ref{fig_confinfA}, right panel, in comparison with the result obtained
by the more accurate formula \eqref{finexpi}. The latter, plotted in the left panel of the figure,  makes use of a small $\Delta \tau$ limit only, without
the further expansion in $1/|\eta|$  of eq \eqref{deflimit}. The resulting profile of the spectrum is very similar in both cases, at least in the  regime $\kappa\le5$, indicating that the  limits of eq \eqref{deflimit} give
trustable results for the spectrum at least at relatively
large scales. It is not difficult to use eq \eqref{simPS} to analytically determine
the position of the dip, finding agreement with 
other works in the literature \cite{Tasinato:2020vdk}.

\begin{figure}
\begin{center}
    \includegraphics[width = 0.58 \textwidth]{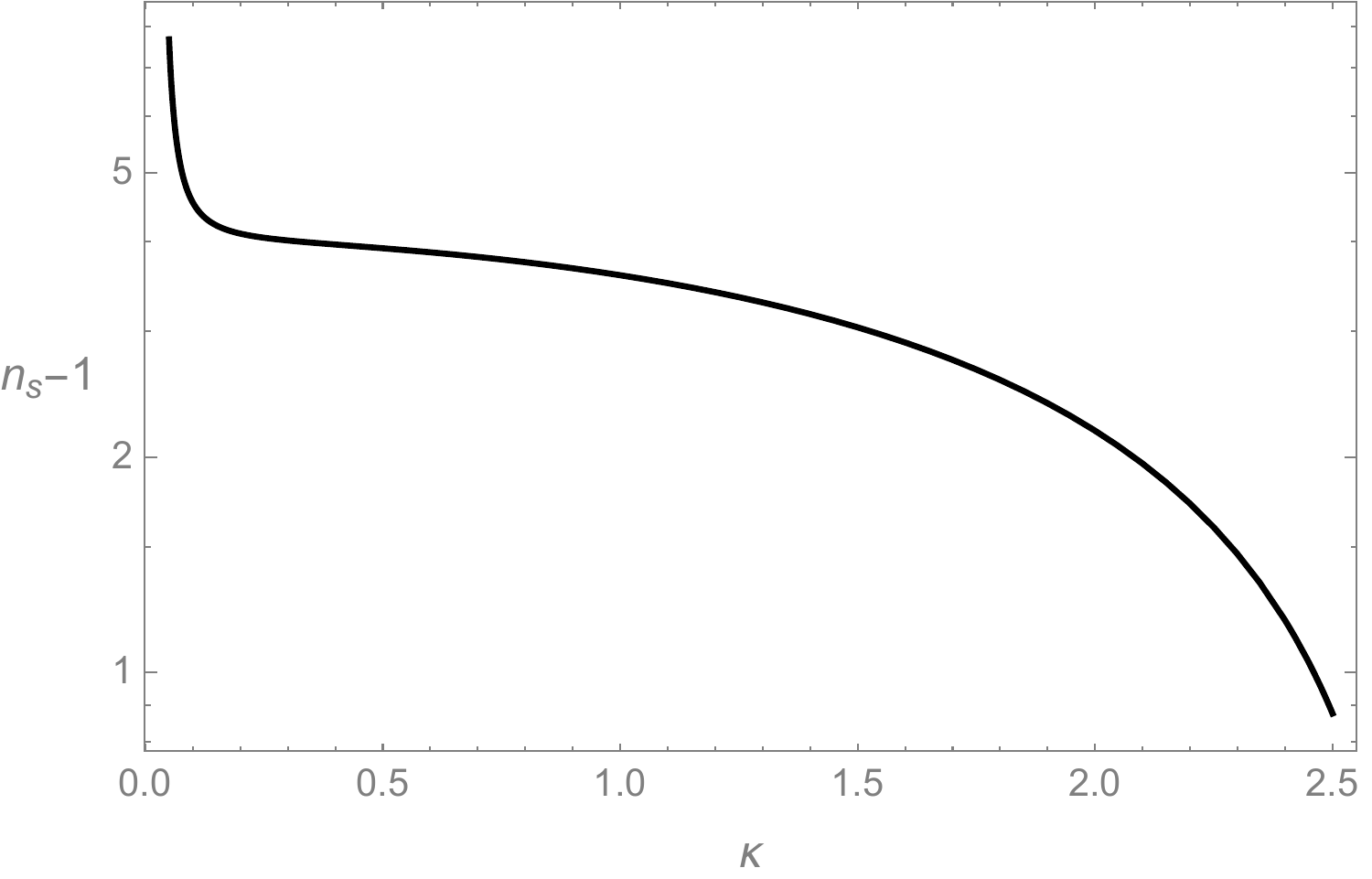}
    \end{center}
 \caption{ \small The spectral index as given in eq \eqref{expNsm1}, choosing $\Pi_0=1250$.}
  \label{fig_confinfBa}
\end{figure}

\smallskip 
It is remarkable to obtain such a simple formula \eqref{simPS} for the scale dependence of the curvature power
spectrum, whose momentum profile shares  features with more realistic PBH models
 discussed in the literature. 
  This formula depends on a single free parameter $\Pi_0$. Besides
 parameterizing  the total enhancement of the spectrum, this  quantity also governs the scale dependence of the spectrum at large scales. Expanding \eqref{simPS} up to 
 $\kappa^2$:
  \be\label{smke}
 \hat \Pi(\kappa)\,=\,1-\frac{4\,\Pi_0}{3} \kappa^2+{\cal O}(\kappa^4)\,,
 \ee
making manifest the role of $\Pi_0$ in controlling the deviations from a flat 
 spectrum.  (See also Starobinsky's scenario \cite{Starobinsky:1992ts} for a specific model leading to an interesting analytical formula for the curvature spectrum beyond slow-roll.)
  We can be more precise 
 and analytically compute the spectral index
associated with eq \eqref{simPS}:
\bea 
\hat n_s(\kappa)-1&\equiv&\frac{d \ln \hat \Pi(\kappa)}{d \ln \kappa}\,,
\\
&=&\frac{2 \,\kappa\, \Pi_0\left[ (1-2 \kappa^2)
\sin{(2 \kappa)}
-2 \kappa \cos{(2 \kappa)}
\right]
}{\kappa^2+4 \kappa \Pi_0 \cos{\kappa}  \left(\kappa \cos{\kappa}
-\sin{\kappa}
\right)+  4 \Pi_0^2  \left(\kappa \cos{\kappa}
-\sin{\kappa}
\right)^2}
\nonumber
\\&&-\frac{\Pi_0^2 \left[4-(4-8 \kappa^2) \cos{(2 \kappa)}
+4 \kappa (\kappa^2-2) \sin{(2 \kappa)}
 \right]}{
 \kappa^2+4 \kappa \Pi_0 \cos{\kappa}  \left(\kappa \cos{\kappa}
-\sin{\kappa}
\right)+  4 \Pi_0^2  \left(\kappa \cos{\kappa}
-\sin{\kappa}
\right)^2
 }\label{expNsm1}\,.
\eea
The rich dependence in  momentum scale of the spectral
index in eq \eqref{expNsm1} reflects
the scale dependence of the spectrum in Fig \ref{fig_confinfA}. We represent it in Fig \ref{fig_confinfBa} for a range
of momenta 
going from the dip position to small scales.
Comparing  Figures \ref{fig_confinfA} and \ref{fig_confinfBa}, 
 we notice that,  after the dip position,  the maximal growth slope
of the spectrum is  $n_s-1\le 4$. This     agrees with the more sophisticated
analysis \cite{Byrnes:2018txb} based on complete expressions for the curvature power spectrum,
outside the large $|\eta|$ limit we consider here.

\section{The three-point function  of curvature fluctuations}
\label{sec_three}

We now apply the previous set-up 
to the study of  three-point function of curvature fluctuations, evaluated at the end of inflation. This quantity controls the non-Gaussianity of curvature
fluctuations in PBH scenarios.
We assume that the slow-roll parameter $\epsilon(\tau)$
remains always small, while $\eta(\tau)$ experiences a  sharp transition between
the slow-roll and non-slow-roll phases, at $\tau=\tau_1$ and $\tau=\tau_2$. The
$n$-point functions of $\zeta$ can be computed using the in-in formalism \cite{Maldacena:2002vr,Chen:2006nt,Seery:2005wm}. Let ${\cal O}(\tau)$
 the operator  one wishes to determine (for us, the  three-point function 
$\langle
 \zeta_{\kappa_1}(\tau_0)
  \zeta_{\kappa_2}(\tau_0)
 \zeta_{\kappa_3}(\tau_0)
\rangle$),  and $ {\cal H}_{\rm int}$ the interaction Hamiltonian.
We map the time evolution
of the operator from the
initial $\big| {\rm in}\rangle$ vacuum up to the time the operator ${\cal O}(\tau)$ is evaluated, and then we map back to the $\big| {\rm in}\rangle$ vacuum again. In formulas: 
$\langle {\rm in} \Big|
{\bar T} e^{-i \int {\cal H}_{\rm int}( \tau') d  \tau'}\,{\cal O}(\tau)\,
{ T} e^{i \int {\cal H}_{\rm int}( \tau') d  \tau'}
 \Big|
{\rm in}  \rangle$. 
In our case, since we focus on sudden transitions, there is a single dominant contribution
to the interaction Hamiltonian \cite{Kristiano:2022maq,Kristiano:2023scm}, which can be extracted~\footnote{The complete third-order Lagrangian density for scalar fluctuations  in single field inflation can be found in eq (3.9) of the original work by Maldacena  \cite{Maldacena:2002vr} (see also eq (35) of \cite{Kristiano:2023scm}). All terms of the Lagrangian are slow-roll suppressed, but there is a single contribution proportional to the time-derivative of $\eta$: $1/2\,\epsilon\,\eta'\,\zeta'\,\zeta^2$. This single term can give a large effect in our context with sudden transitions between slow-roll and non-slow-roll eras.  As done in \cite{Kristiano:2022maq,Kristiano:2023scm}, we then focus on this contribution to the third order action, leading to the interaction Hamiltonian in eq \eqref{intham}.} from the third-order action of
perturbations in single field inflation \cite{Maldacena:2002vr}:
\be
\label{intham}
{\cal H}_{\rm int}\,=\,-\frac12 \int d^3 x\,a^2(\tau) \epsilon(\tau)\,\eta'(\tau)\,\zeta^2(\tau,\vec x)
\,\zeta'(\tau,\vec x)\,.
\ee
We assume that $|\eta|$ is negligible during slow-roll evolution ($\tau<\tau_1$ and
$\tau_2<\tau<\tau_0$) 
while it is large during the intermediate NSR  phase, $\tau_1\le\tau\le\tau_2$.
We adopt a sharp-transition Ansatz \cite{Kristiano:2023scm} for the time-derivative of $\eta(\tau)$
\be
\eta'(\tau)\,=\,\Delta \eta \left[ -\delta(\tau-\tau_1)
+\delta(\tau-\tau_2)
\right]\,.
\label{ansetp}
\ee
where the  times $\tau_{1,2}$ correspond to the onset and end
of the NSR phase during inflation. Sudden transitions with jumps
in the first derivatives of the inflation scalar profile and the parameter
$\eta$ can be explicitly realised in scenarios as the Starobinsky model \cite{Starobinsky:1992ts},
with a piece-wise linear potential characterised by abrupt changes in its slope. 

\smallskip

Soon we will discuss  a criterium to select the constant $\Delta \eta$.  But first, we
apply the aforementioned in-in approach with the interaction
Hamiltonian \eqref{intham}, and eq \eqref{ansetp}. 
 The curvature three-point
function, evaluated at the end of inflation $\tau_0(=0)$, results \cite{Kristiano:2023scm}
\bea
&&\langle  \zeta_{\kappa_1}(\tau_0)  \zeta_{\kappa_2}(\tau_0)
 \zeta_{\kappa_3}(\tau_0)
 \rangle'
 =
 \nonumber
 \\
 &&-2 \Delta \eta \Big(\epsilon(\tau_2) a^2(\tau_2)
 \,{\rm Im}\left[
 \left(  \zeta_{\kappa_1}(\tau_0)   \zeta^{*}_{\kappa_1} (\tau_2) \right)
  \left(  \zeta_{\kappa_2}(\tau_0)   \zeta^{*}_{\kappa_2} (\tau_2) \right)
 \left(  \zeta_{\kappa_3}(\tau_0)  \partial_{
 \tau_2} \zeta^{*}_{\kappa_3} (\tau_2) \right)
   \right]-(\tau_2\to\tau_1) \Big)
    \nonumber
 \\
 &&+{\rm perms\,.}
 \label{bisopun}
\eea
where recall that the prime means that we understand the momentum-conserving
delta functions.
In the squeezed limit, eq \eqref{bisopun} reduces to
\bea
&&\lim_{\kappa_1\to0;\,\kappa_2\simeq \kappa_3}\langle \zeta_{\kappa_1}(\tau_0) \zeta_{\kappa_2}(\tau_0)\zeta_{\kappa_3}(\tau_0)
\rangle' \nonumber
\\
&&=\,-\,4 \Delta \eta\,\epsilon(\tau_2) a^2(\tau_2)\,|\zeta_{\kappa_1}(\tau_0)|^2\,|\zeta_{\kappa_2}(\tau_0)|^2
\nonumber
\\
&&\times
\left\{
{\rm Im}\left[ \frac{\zeta_{\kappa_2}^2(\tau_0)}{|\zeta_{\kappa_2}(\tau_0)|^2}
\zeta_{\kappa_2}^*(\tau_2)(\zeta'_{\kappa_2}(\tau_2))^*
\right]-\frac{\epsilon(\tau_1) a^2(\tau_1)}{\epsilon(\tau_2) a^2(\tau_2)}
{\rm Im}\left[ \frac{\zeta_{\kappa_2}^2(\tau_0)}{|\zeta_{\kappa_2}(\tau_0)|^2}
\zeta_{\kappa_2}^*(\tau_1)(\zeta'_{\kappa_2}(\tau_1))^*
\right]
\right\}\,.
\nonumber
\\
\label{sqztp}
\eea
The squeezed limit refers to modes with very small momenta $\kappa_1$
which leave the horizon much earlier than the onset of the non-slow-roll (NSR) phase.
When selecting large-scale modes with $\kappa_2$ small, also far from
the NSR epoch, we do expect that the standard Maldacena consistency relation \cite{Maldacena:2002vr} holds. Namely
\bea
\label{maldacond}
&&\lim_{\kappa_1\to0;\,\kappa_2\simeq \kappa_3}\langle \zeta_{\kappa_1}(\tau_0) \zeta_{\kappa_2}(\tau_0)\zeta_{\kappa_3}(\tau_0)
\rangle'
\,=\,-(n_s(\kappa_2)-1)\,|\zeta_{\kappa_1}(\tau_0)|^2\,|\zeta_{\kappa_2}(\tau_0)|^2\,.
\eea

\begin{figure}
\begin{center}
    \includegraphics[width = 0.45 \textwidth]{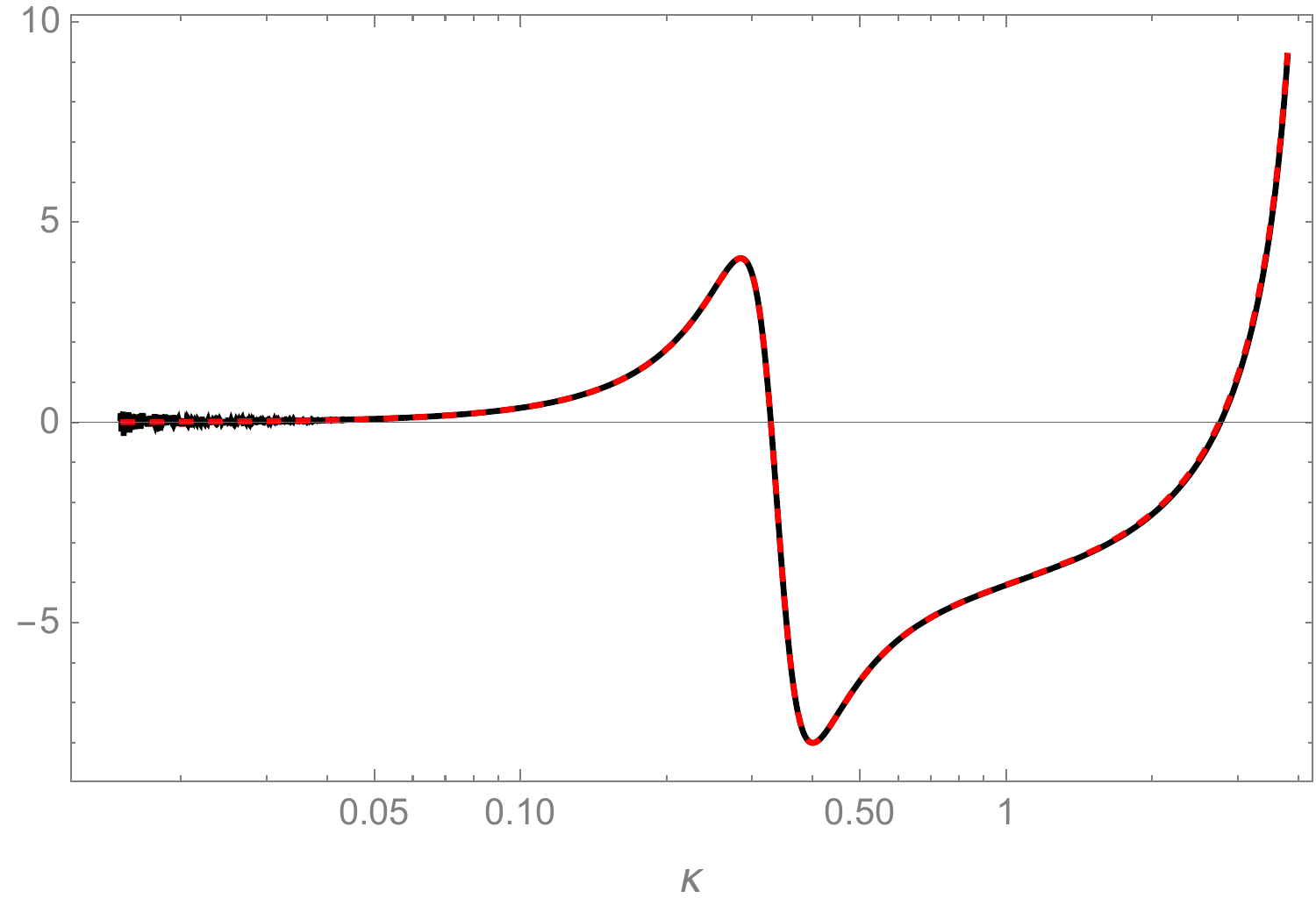}
        \includegraphics[width = 0.52 \textwidth]{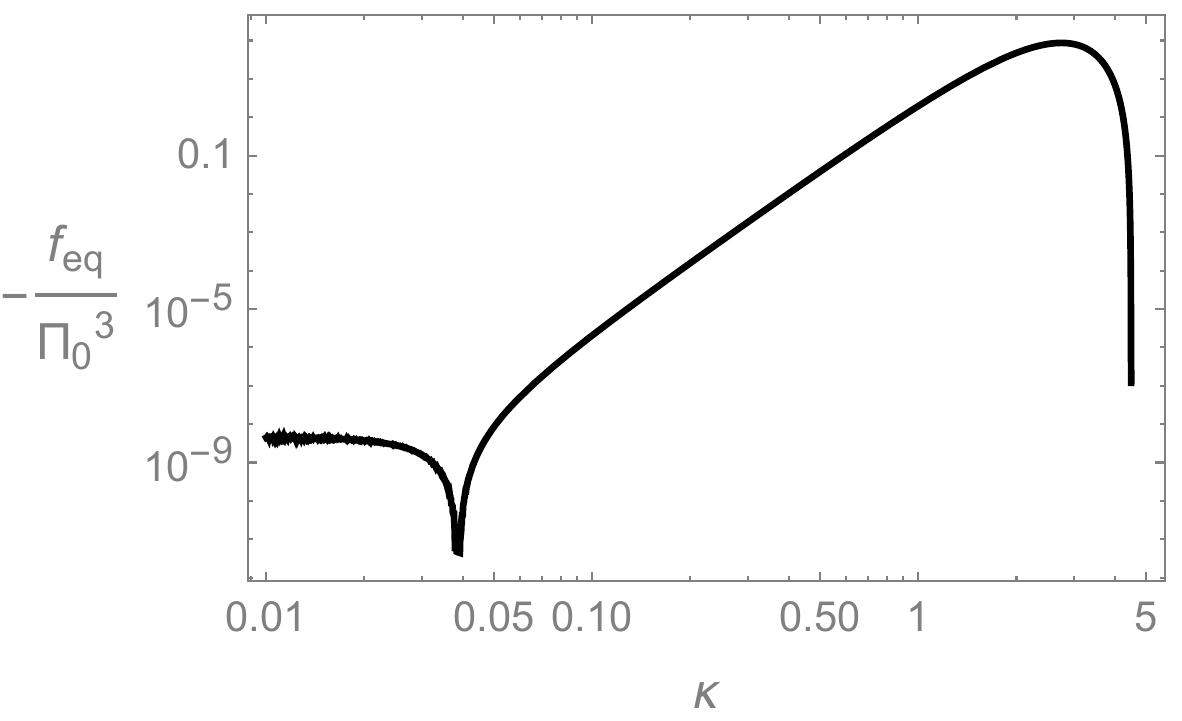}
    \end{center}
 \caption{ \small {\bf  Left panel} Check of Maldacena consistency relation for the squeezed limit of the three point function.  Black line: the
 quantity $1-n_s$. Red dashed line: the squeezed limit of the three-point function of eq \eqref{sqztp} (we omit  the factors $|\zeta_{\kappa_1}(\tau_0)|^2\,|\zeta_{\kappa_2}(\tau_0)|^2$). 
 We use the  mode functions in eqs \eqref{modeNSR}, and choose the values
  $|\eta|=10^{4.1}$, $\Delta \tau=0.002$.
  {\bf Right panel:} Plot of the scale dependence of the equilateral three-point function, the quantity $-f_{\rm eq}/\Pi_{0}^3$, as defined in the main text, eq \eqref{eqfnl}. The profile
  is remarkably similar to the power spectrum of Fig \ref{fig_confinfA}.
 }
  \label{fig_confinfB}
\end{figure}

We substitute our expressions for the mode functions
in eq \eqref{modeNSR},  and take the small $\kappa_2$ limit.   Using the results of 
Section \ref{sec_power} for computing 
 the spectral index,  the two expressions \eqref{sqztp} and \eqref{maldacond}
match once we select a certain value for the  parameter $\Delta \eta$
which enters in the Ansatz \eqref{ansetp}. Neglecting contributions that vanish 
 in the large-$|\eta|$
limit,  
   we find
the requirement
\be
\label{solde}
\Delta \eta\,=\,\frac{|\eta|}{(1+\Pi_0)}+\frac{\Pi_0 (12+34 \Pi_0+25 \Pi_0^2)}{2 (1+\Pi_0)^2 (1+2 \Pi_0)}\,,
\ee
as well as the expected
condition  \footnote{In fact, we are working in a regime of large
$|\eta|$ and very small $\Delta \tau$, as dictated by relations
\eqref{deflimit}. Hence, we obtain
\bea
\epsilon(\tau_1)\,a^2(\tau_1)&=&\epsilon(\tau_2)\,a^2(\tau_2)\left(1+\tau_1\,\frac{\epsilon'(\tau_1)}{\epsilon(\tau_1)}
\,\Delta \tau\right)\left(1+2 \tau_1\,\frac{a'(\tau_1)}{a(\tau_1)}
\,\Delta \tau\right)\,,
\nonumber
\\
&=&\epsilon(\tau_2)\,a^2(\tau_2)\left(1+\tau_1\,a(\tau_1) H(\tau_1)\left(\eta+2 \right)
\,\Delta \tau \right)\,,
\nonumber
\\
&=&\epsilon(\tau_2)\,a^2(\tau_2)\left(1-\left(\eta+2 \right)
\,\Delta \tau \right)\,,
\nonumber
\\
&=&\epsilon(\tau_2)\,a^2(\tau_2)\left(1+2 \Pi_0\right)\,,
\nonumber
\eea
in agreement with condition \eqref{solde2}.}
%, the requirement to we find
\bea
\frac{\epsilon(\tau_1) a^2(\tau_1)}{\epsilon(\tau_2) a^2(\tau_2)}
\,=\,1+2 \Pi_0\,.
\label{solde2}
\eea
%as expected in this regime \footnote{In fact, \dots}. 
Interestingly, although 
the constant $\Delta \eta$ has been fixed to satisfy Maldacena condition
in the small-$\kappa_2$ limit,  the resulting expression \eqref{sqztp} for the squeezed three-point function
that matches well with single-field Maldacena consistency relation also for larger scales: see
 Fig \ref{fig_confinfB}, left panel, which is also
 in agreement with \cite{Ozsoy:2021qrg,Ozsoy:2021pws}. 
 The resulting squeezed non-Gaussianity is strongly scale-dependent \cite{Byrnes:2009pe,Byrnes:2010ft}.
 
 \bigskip

  The squeezed limit of the three-point function, as in eq \eqref{sqztp}, is
 not the only interesting configuration. From the complete expression
 for the three-point function, eq \eqref{bisopun}, we can also consider other shapes. 
 For example, let us consider the equilateral limit $\kappa_i=\kappa$ for $i=1,2,3$.
 In Fig \ref{fig_confinfB}, right panel, we represent the value for the three-point function
 as a function of the dimensionless scale $\kappa$, divided by the square
 of the large-scale power spectrum, eq \eqref{lsspec} (we further divide it by $\Pi_0^3$). 
  Namely,
  \be
 \label{eqfnl}
\frac{ f_{\rm eq}(\kappa)}{\Pi_0^3}\equiv\frac{\langle 
\zeta_{\kappa}(\tau_0)  \zeta_{\kappa}(\tau_0)
 \zeta_{\kappa}(\tau_0)
\rangle'}{\Pi_0^3\,{\cal P}_0^2}\,.
 \ee 
  This
 quantity aims to capture the scale-dependence of the non-Gaussian equilateral limit
 \cite{Chen:2005fe}, 
 analogously to the scale-dependent part of the power spectrum of eq \eqref{lsspec}.
 Remarkably, 
 the profile of the scale-dependence for the equilateral shape (changing its overall sign) is  similar
 to the profile  of the scale-dependent power spectrum: compare
  Fig \ref{fig_confinfA} with Fig \ref{fig_confinfB}, right panel. It would
 be interesting to find a physical reason for this result.

\bigskip

The  non-Gaussianity of curvature
fluctuations in PBH scenarios is an important
observable with several phenomenological ramifications for PBH formation \cite{Byrnes:2012yx,Young:2013oia,Passaglia:2018ixg,Biagetti:2021eep,Atal:2018neu}.
We refer the reader to \cite{Taoso:2021uvl} for a  recent comprehensive analysis, and further references
therein.

%%%%%%%%%%%%%%%%%%%%%%%%%%%%%%%%%%%%%%%%
%%%%%%%%%%%%%%%%%%%%%%%%%%%%%%%%%%%%%%%%
\section{Loop corrections  }

In this section we apply the previous tools to study loop contributions
to the inflationary power spectrum \cite{Weinberg:2005vy,Sloth:2006az,Seery:2007wf,Senatore:2009cf,Bartolo:2010bu}. In developing our arguments, we closely follow the 
clear technical discussion of \cite{Kristiano:2023scm}, but we make use of our large $|\eta|$ expansion, and the
corresponding  solutions for the mode functions discussed in Section \ref{sec_system}.   
 We are especially
interested in examining the physical implications of large loop corrections in our approach,
and the role of the scale dependence of the spectrum. Moreover, we discuss a proposal
  to  adsorb quadratic ultraviolet  divergences into the available bare parameters
  in a large $|\eta|$ limit, at least
  at large scales relevant for CMB physics. We are left with log-enhanced, infrared
  effects whose size is small at large scales. 
   This is an important 
step in order to clarify the relation between loops and physically measurable quantities.

\bigskip

The interaction
Hamiltonian that we consider is given in eq \eqref{intham}; as in Section \ref{sec_three},  we focus 
on a sharp transition between slow-roll regimes and an intermediate non-slow-roll
regime for $\tau_1\le\tau\le\tau_2$. 
 We consider for definiteness
 the two-point function of curvature fluctuations in momentum space, 
evaluated at the  scale $p$ (dimensionless in the sense that the momentum is multiplied by $-\tau_1$, as in eq \eqref{dimkd}). 
The corresponding 1-loop 
contributions can be found utilising the in-in formalism.
 We follow \cite{Kristiano:2023scm}: loop corrections are conveniently decomposed as 
 \bea
\langle \zeta_{ p}(\tau_0)\zeta^*_{ p}(\tau_0)
\rangle_{\rm loop}\,=\,
\langle \zeta_{ p}(\tau_0)\zeta^*_{ p}(\tau_0)
\rangle_{(1,1)}+
2\,{\rm Re}\left[
\langle \zeta_{ p}(\tau_0)\zeta^*_{ p}(\tau_0)
\rangle_{(2,0)}
\right]\,,
\eea
where each term correspond to a different contribution in the expansion of the two-point correlator in the in-in formalism \cite{Kristiano:2023scm}. They read
\bea
\langle \zeta_{ p}(\tau_0)\zeta^*_{ p}(\tau_0)
\rangle_{(1,1)}&=&\frac14\,\int_{-\infty}^{\tau_0}\,d \tau_a\,a^2(\tau_a)\,\epsilon(\tau_a)\,\eta'(\tau_a)
\int_{-\infty}^{\tau_0}\,d \tau_b\,a^2(\tau_b)\,\epsilon(\tau_b)\,\eta'(\tau_b)
\nonumber
\\
&&\times \int \Pi_{i=1}^6\, \frac{d^3 k_i}{(2 \pi)^3} \,\delta^{3}(\vec k_1+ \vec k_2+\vec k_3)%\,\delta^{3}(\vec k_1+ \vec k_2+\vec k_3)
\,\delta^{3}(\vec k_4+ \vec k_5+\vec k_6)
\nonumber
\\
&& \times \langle
\zeta'_{\vec k_1}(\tau_a)
\zeta_{\vec k_2}(\tau_a)
\zeta_{\vec k_3}(\tau_a)
\zeta_{\vec p}(\tau_0)
\zeta_{-\vec p}(\tau_0)
\zeta'_{\vec k_4}(\tau_b)
\zeta_{\vec k_5}(\tau_b)
\zeta_{\vec k_6}(\tau_b)
 \rangle\,,
\eea
and
\bea
\langle 
\zeta_{ p}(\tau_0)\zeta^*_{ p}(\tau_0)
%\zeta_{\vec p}(\tau_0)\zeta_{-\vec p}(\tau_0)
\rangle_{(2,0)}&=&-\frac14\,\int_{-\infty}^{\tau_0}\,d \tau_a\,a^2(\tau_a)\,\epsilon(\tau_a)\,\eta'(\tau_a)
\int_{-\infty}^{\tau_0}\,d \tau_b\,a^2(\tau_b)\,\epsilon(\tau_b)\,\eta'(\tau_b)
\nonumber
\\
&&\times \int \Pi_{i=1}^6\, \frac{d^3 k_i}{(2 \pi)^3} \,\delta^{3}(\vec k_1+ \vec k_2+\vec k_3)%\,\delta^{3}(\vec k_1+ \vec k_2+\vec k_3)
\,\delta^{3}(\vec k_4+ \vec k_5+\vec k_6)
\nonumber
\\
&& \times \langle
\zeta_{\vec p}(\tau_0)
\zeta_{-\vec p}(\tau_0)
\zeta'_{\vec k_1}(\tau_a)
\zeta_{\vec k_2}(\tau_a)
\zeta_{\vec k_3}(\tau_a)
\zeta'_{\vec k_4}(\tau_b)
\zeta_{\vec k_5}(\tau_b)
\zeta_{\vec k_6}(\tau_b)
 \rangle\,.
 \label{genlop1}
\eea

\smallskip

From now on, to simplify the calculations we focus on a large-scale regime where the size of the external momentum
$p$ is much smaller than the momenta $k_i$ over which we integrate \cite{Kristiano:2022maq,Kristiano:2023scm}. This allows to simplify formulas substituing $k-p\simeq k$, and permits us to obtain analytic results. We will discuss in due course the limitations we should impose on $p$  for satisfying this condition, and their physical implications.

\bigskip

Substituting our Ansatz \eqref{ansetp}  in the case  a sharp transition at the times $\tau_1$
and $\tau_2$ between slow-roll and non-slow-roll phases, the result acquires
the following structure \cite{Kristiano:2023scm} 
\bea
\langle 
\zeta_{ p}(\tau_0)\zeta^*_{ p}(\tau_0)
\rangle_{\rm loop}&=& \left( 2\epsilon(\tau_2) a^2(\tau_2)\right)^2 \Delta \eta^2\,|\zeta_{\vec p}(\tau_0)|^2\,
\nonumber
\\
&\times&
\int \frac{d^3 k}{(2 \pi)^3}
 \Big[|\zeta_{\vec k}(\tau_2)|^2 \,{\rm Im}(\zeta_p(\tau_2) \zeta'^*_p(\tau_2))\,{\rm Im}(\zeta_k(\tau_2) \zeta'^*_k(\tau_2))
\nonumber
\\
&&-4\frac{ \epsilon(\tau_1) a^2(\tau_1)}{\epsilon(\tau_2) a^2(\tau_2)}\,{\rm Im}(\zeta_p(\tau_0) \zeta^*_p(\tau_2))\,{\rm Im}(\zeta_k'(\tau_2) \zeta_k(\tau_2) 
\zeta^*_k(\tau_1)\zeta'^*_k(\tau_1)
)
\nonumber
\\
&&-2\frac{ \epsilon(\tau_1) a^2(\tau_1)}{\epsilon(\tau_2) a^2(\tau_2)}\,{\rm Im}(\zeta_p(\tau_2) \zeta'^*_p(\tau_2))\,{\rm Im}(\zeta_k^2(\tau_2) \zeta_k^*(\tau_1) 
\zeta_k'^*(\tau_1)
)\nonumber
\\
&&+\frac{ \epsilon^2(\tau_1) a^4(\tau_1)}{\epsilon^2(\tau_2) a^4(\tau_2)}\,|\zeta_{\vec k}(\tau_1)|^2\,{\rm Im}(\zeta_p(\tau_1) \zeta'^*_p(\tau_1))\,{\rm Im}(
\zeta_k(\tau_1) \zeta_k'^*(\tau_1) 
)
\Big]\,.
\label{genlo2}
\eea
Since the integrand functions are rotationally invariant, the three-dimensional integrals over internal
momenta can be decomposed into 
   integrals over the real line  as
 \bea
 \int \frac{d^3 k}{(2 \pi)^3}\left(\dots\right)&=&\int_{\Lambda_{\rm IR}/{|\eta|}^{1/2}}^{\mu/|\eta|^{1/2}} \frac{k^2\,d k}{2 \pi^2}\left(\dots\right)+
 \int^{\Lambda_{\rm UV}/{|\eta|}^{1/2}}_{\mu/|\eta|^{1/2}} \frac{k^2\,d k}{2 \pi^2}\left(\dots\right)
 \label{strlop1}
 \eea
 with $\Lambda_{\rm IR}$ and $\Lambda_{\rm UV}$ corresponding to a very small infrared (IR) and a very large  ultraviolet (UV) 
 cut-off \footnote{From now on, our approach is a different  from  \cite{Kristiano:2022maq}, where
  $\Lambda_{\rm IR, UV}$ are   scales of modes  leaving the horizon at the start and  end of the NSR era. In our case, being the NSR epoch very short, we do not make this
  identification. }.  They are  dimensionless
 quantities, obtained multiplying physical momentum  scales with $|\tau_1|$,
 as in eq \eqref{dimkd}.
 For convenience, as a technical device we rescale the extrema of integration by  $1/|\eta|^{1/2}$,  to simplify our results in a large $|\eta|$ limit. The intermediate dimensionless scale $\mu$ is introduced in order to physically separate the loop corrections in an IR part (the first
 integral in eq \eqref{strlop1}) and a UV part (the second integral).  We can think of $\mu\sim1$ as a scale where NSR effects take place.  This separation will be essential for our arguments.

\smallskip

We  decompose the resulting power spectrum at the end of inflation $\tau_0=0$ as 
a tree level and a loop part
\bea
{\cal P}_{\rm tot}(p)&=&\frac{p^3}{2 \pi^2 (-\tau_1)^3}\,\langle \zeta_{ p}(\tau_0)\zeta^*_{ p}(\tau_0)\rangle
\,=\,{\cal P}_0\,\hat \Pi(p)\left[1+{L}^{\rm IR}_{\rm loop}+{L}^{\rm UV}_{\rm loop}\right]
\label{dec_ps}
\,,
\eea
with ${\cal P}_0$ the amplitude of the large scale tree-level power spectrum as in eq \eqref{lsspec}, and  
  $\hat \Pi(p)$ is the momentum-dependent function of eq \eqref{simPS}, controlling the scale-dependent ratio 
between small-scale and large-scale spectra. 
 Eq \eqref{dec_ps}
contains the quantity
 ${L}_{\rm loop}$,
 the loop contribution \eqref{genlo2}, with the momentum
integrals decomposed as in eq \eqref{strlop1}. We collect as  an overall factor the  
momentum dependent quantity  ${\cal P}_0\,\hat{\Pi}(p)$.

\bigskip

We substitute in the general formulas \eqref{strlop1} our
mode functions \eqref{modeNSR}. We analytically perform both the IR and the UV integrals, which are
 much simplified in the large $|\eta|$ regime of  eq \eqref{deflimit}, which keeps
$\Pi_0$ fixed.  
At leading order in $1/|\eta|$, the dominant   contribution to the IR piece of the loop correction results 
\bea
\label{anloopr1a}
{L}^{\rm IR}_{\rm loop}&=&-p^2\,\frac{{\cal P}_0}{3}\,\frac{\Pi_0^4}{(1+\Pi_0)^2(1+2 \Pi_0)}\,\ln{\left(\frac{\mu}{\Lambda_{\rm IR}} \right)}\,,
\eea
where we include only the log-enhanced part. Notice that the IR contribution is proportional to $p^2$, hence it is suppressed at large scales. As in the rest of the work, the quantity $p$ is the dimensionless momentum scale obtained dividing the physical momentum by the pivot scale $k_\star$ (see eq \eqref{dimkd}). 
We neglect power-law quadratic pieces
depending on the small quantity $\Lambda_{\rm IR}$, and on $\mu$ which, being of order 
one, is suppressed with respect to the logarithm in eq \eqref{anloopr1a},
in case of a large ratio ${\mu}/{\Lambda_{\rm IR}}$.
 This
  IR contribution can be interpreted
  as
   a secular effects caused by modes crossing the
 horizon from the onset of inflation until around   the epoch of NSR, controlled
 respectively by 
 the scales $\Lambda_{\rm IR}$ and $\mu$.  
  IR  contributions are typically characterized by large logarithms,
 whose effects   might contribute to observable quantities, if inflation lasts long. In
 our case, we can estimate $\ln{\left({\mu}/{\Lambda_{\rm IR}} \right)} \sim \ln\left[{a(\tau_1)/a(\tau_{\rm start})}\right]$). Hence,  we can expect the logarithm to be  of order say $10^2$.  
  See e.g. the clear 
 discussion in \cite{Sloth:2006az}.

The dominant contribution to the UV integral is a quadratic divergence in $\Lambda_{\rm UV}$. We write the result only including the contribution quadratic in the UV cut-off
\bea
{L}^{\rm UV}_{\rm loop}&=&
 -\frac{{\cal P}_0\,\Pi_0\,
 \Lambda_{\rm UV}^2 %- \Lambda_{\rm IR}^2\right)
 }{(1+\Pi_0)}\left(\frac56+  \frac{3 j_1(p)-p}{3 \,p} \right)\,,
\label{anloopr1b}
 \eea
 and we neglect subleading contributions.  As the IR contribution, also the UV part depend quadratically on a scale, this time the 
cut-off scale $\Lambda_{\rm UV}$. As explained after eq \eqref{strlop1}, the cut-off scales are again  dimensionless quantities, obtained
dividing physical cut-off scales by a pivot scale. 
 The spherical Bessel function $j_1(p)$ defined in eq \eqref{defj1}.
  The UV part contains the effects of small-scale modes, which remain in a thermal vacuum 
 within a subhorizon regime during the first phase of inflation, until   the short NSR phase occurs. These modes should
 not   participate to the dynamics of the NSR era during inflation, and the associated UV divergences are  expected to be adsorbed into appropriate, physically measurable
 quantities
 % appropriate renormalization procedure
 (see e.g. \cite{Boyanovsky:2005sh} for a detailed analysis within slow-roll models). 

\smallskip
We adopt this viewpoint, and assume that the contributions of ${L}^{\rm UV}_{\rm loop}$
are adsorbed into the available parameters by means of an appropriate
renormalization procedure. We discuss in Appendix \ref{app_reno} a way to do so. 
We are left with the log-enhanced loop contributions
of eq \eqref{anloopr1a}.
All our results are  derived under the approximation stated after eq \eqref{genlop1}: to analytically compute the integrals, we make the hypothesis that the momentum $p$ at which 
 the two-point function \eqref{dec_ps} is computed is  well {\it  smaller} than the momentum scales over which integrate, $i.e.$ the  lower extremum of the integral
 \be
\label{pcon1}
 p^2
 \,
 \le 
 \,\frac{ \Lambda_{\rm IR}^2}{|\eta|}\,.
 \ee 
 Since we are working at leading order in a $1/|\eta|$ expansion, the previous condition informs us that we should only focus on the very first terms in a momentum expansion of our formulas. Using the expression \eqref{simPS}, 
 we consider eq \eqref{dec_ps} up to second order in an expansion in momentum $p$,
 including the IR loop contributions:
 \bea
 \label{dec_ps2a}
 {\cal P}_{\rm tot}(p)&=&{\cal P}_0%+{\cal P}_0^2\,{L}_{\rm loop}^{(a)}
 -\frac{4{\cal P}_0\,\Pi_0}{3} \left[1+
 \,\frac{{\cal P}_0}{4}\,\frac{\Pi_0^3}{(1+\Pi_0)^2(1+2 \Pi_0)}\,\ln{\left(\frac{\mu}{\Lambda_{\rm IR}} \right)}
  \right]\,{p^2}+{\cal O}(p^4)\,.
 \eea
 Hence, the log-enhanced IR loop only gives a contribution to the quadratic
 term in the expansion. Its size is small, being suppressed by a factor ${\cal P}_0\simeq 10^{-9}$ with respect to the tree level term, so even a large logarithm is 
 unable to give large effects.  The coefficients depending on $\Pi_0$ give
 order one effects, in the limit of $\Pi_0$ large. 
 
 \smallskip
 
 We conclude this Section  with a comparison between our results and some recent papers appeared in the literature on 
 this subject. The works \cite{Kristiano:2022maq,Kristiano:2023scm} focus  on the ultra-slow-roll case $|\eta|=6$,
 assuming  sudden transitions between slow-roll and non-slow-roll epochs. In this specific case, loop corrections give dangerously 
 large contributions to correlators evaluated at large, CMB scales. Recently, the works \cite{Motohashi:2023syh,Franciolini:2023lgy}
 shown that for different values of $|\eta|$ in the non-slow-roll epoch loop corrections can be reduced, and do not necessarily spoil
 CMB predictions. Our large $|\eta|$ approach lies on this category of scenarios: it explores systems with $|\eta|\neq 6$, confirming that
 regions in parameter space with   $|\eta|\gg 6$ lead to loop corrections suppressed by powers of momentum $p$ (see eq. \eqref{dec_ps2a}), negligible at large scales.  Another possibility to explore is relaxing the assumption of sudden
 transition between slow-roll and ultra-slow-roll phases. This idea has been carried on in \cite{Riotto:2023gpm,Firouzjahi:2023aum,Firouzjahi:2023ahg }, showing that loop effects can be considerably reduced
  when adopting a smooth transition between different epochs. The same remains true including higher order effects
 associated with quartic interactions \cite{Firouzjahi:2023aum}. It would be very interesting to further explore these topics 
 in the context of our large $|\eta|$ approach, also studying the effects of loops at large values of momenta  (as done for
 example in \cite{Franciolini:2023lgy}), and carrying on a renormalization procedure using for example the  techniques   pursued by \cite{Choudhury:2023hvf} in a related context. Moreover, it would be interesting to study higher loops, and consequences of higher order interactions making use of  a large $|\eta|$ expansion.
   We leave these investigations
  for  future works.

% \smallskip
 
% ------------------------------------------------------------------------------
 
% The main difference with respect to \cite{Kristiano:2022maq}, who
 %focussed on $|\eta|=6$, is that the physical
  %loop effects are suppressed by a factor $p^2$, rendering them small at large CMB scales $p\ll1$. 
%It would be interesting to pursue our large-$|\eta|$ program further,  estimating the loop corrections
%also at smaller  scales of $p$ of order one,  understanding whether the associated divergences can be adsorbed
%into measurable quantities. A computation  of loop effects valid for larger $p$  was recently carried out in  the interesting work \cite{Franciolini:2023lgy}
 %using a numerical procedure, in a framework similar to the one of \cite{Kristiano:2022maq}. 
 %Probably,  also in the  large-$|\eta|$ set-up we discuss here, a proper computation of the associated loop integrals need to be handled by means of a numerical approach. It would be important to then 
% investigate at what extend divergences  renormalize the tree-level spectrum, for example
% using more sophisticated techniques as the ones pursued by \cite{Choudhury:2023hvf} 

%%%%%%%%%%%%%%%%%%%%%%%%%%%%
%%%%%%%%%%%%%%%%%%%%%%%%%%%%
%%%%%%%%%%%%%%%%%%%%%%%%%%%%
%%%%%%%%%%%%%%%%%%%%%%%%%%%%
%%%%%%%%%%%%%%%%%%%%%%%%%%%%

\subsection*{Acknowledgments}
It is a pleasure to thank Maria Mylova, Ogan \"Ozsoy, and Ivonne Zavala for useful input. GT is partially funded by the STFC grant ST/T000813/1. For the purpose of open access, the author has applied a Creative Commons Attribution licence to any Author Accepted Manuscript version arising.

\begin{appendix}
\section{Curvature perturbations and the NSR regime}
\label{AppA}

In this technical Appendix, we briefly review  the results 
 developed in \cite{Tasinato:2020vdk} to determine analytic
solutions for inflationary mode functions during  non-slow-roll
regimes, referring the reader to \cite{Tasinato:2020vdk} for more details. Starting from the quadratic action \eqref{quadac} of curvature perturbations,
it is convenient to introduce a Mukhanov-Sasaki variable $v_k (\tau)\,=\,\zeta_k(\tau)/z(\tau)$, satisfying the equation
\be
\label{mseq}
v_k''(\tau)+\left[k^2-\frac{z''(\tau)}{z(\tau)}
\right]\,v_k(\tau)\,=\,0\,,
\ee
in momentum space. In our case, the inflationary evolution for $\tau\le\tau_0$
undergoes different phases. We have
an initial slow-roll phase for $\tau\le\tau_1$, where both the slow-roll parameters
$\epsilon(\tau)$ and $\eta(\tau)$ are very small. We can
approximate this as pure de Sitter phase.
Then, for $\tau_1\le\tau\le \tau_2$
we have non-slow-roll evolution where $\epsilon(\tau)$
keeps small, while $\eta(\tau)$ is negative but potentially large
in size. We denote with $\epsilon_1$ and $\eta$ the values of the slow-roll
parameters evaluated at $\tau\to\tau_1^+$.  Finally, a slow-roll phase $\tau_2<\tau\le\tau_0$
where the slow-roll parameters return to very small values. Again, we 
approximate this last phase to pure de Sitter.  We
assume that the pump field $z(\tau)$ is continuous at the transitions.

 In the de Sitter limit, while $z''(\tau)/z(\tau)\,=\,2/\tau^2$ for  $\tau\le\tau_1$ and  $\tau_2<\tau\le\tau_0$, the time-profile for this quantity can be richer. As in \cite{Tasinato:2020vdk},
 we adopt an Ansatz
 \bea
 v_k(\tau)&=&-\frac{i\,H_0\,z(\tau)\,e^{-i k \tau}}{2\sqrt{\epsilon_1\,k^3} }
\,{\cal C}_1(k) \left[ 1+i k \tau +(i k \tau_1)^2 A_{(2)} (\tau)+(i k \tau_1)^3 A_{(3)} (\tau)  \right]
\nonumber
 \\
 &&-\frac{i\,H_0\,z(\tau)\,e^{i k \tau}}{2\sqrt{\epsilon_1\,k^3} }
\,{\cal C}_2(k) \left[ 1-i k \tau +(-i k \tau_1)^2 A_{(2)} (\tau)+(-i k \tau_1)^3 A_{(3)} (\tau)  \right]
\nonumber
\\
\label{ansfu1}
 \eea
 for the Mukhanov-Sasaki mode function. 
 
 For $\tau<\tau_1$, the mode equation is the same as in a standard slow-roll era:
 in order to match with the Bunch-Davies vacuum, we select $A_{(n)}=0$ for
 $n\ge2$, as well as ${\cal C}_2=0$ and ${\cal C}_1=1$ in eq \eqref{ansfu1}.    
 For $\tau_1\le \tau \le \tau_2$, we can use the Ansatz  \eqref{ansfu1}
 in the evolution equation \eqref{mseq}, and solve the equation order by order in powers of 
 $(k \tau_1)$: see \cite{Tasinato:2020vdk}. At each order $n$ in $(k \tau_1)^n$, the equation can be solved at leading
 order in an expansion  in the parameter $\Delta \tau$ of eq \eqref{defDT}  controlling the duration
 of the non-slow-roll era. 
 For each $n$, the result depends on powers of the quantity $d \ln{[z^2(\tau)/a^2(\tau)]}/d \ln \tau$, evaluated at time $\tau_1^+$ at the onset of the NSR era.  
This quantity was dubbed $\alpha$ in \cite{Tasinato:2020vdk}: in the present instance, within single
field inflation with canonical kinetic terms and in a pure de Sitter limit,
it 
 corresponds to the quantity $-\eta$ (we 
use
the definitions \eqref{defsr1}). 
 After computing each quantity  $A_{(n)}$,  
   the resulting series in eq \eqref{ansfu1} can be resummed analytically
 in terms of exponentials. 
 The result of the resummation
    is \cite{Tasinato:2020vdk} 
 \bea
 v_k(\tau)&=&-\frac{i\,H_0\,z(\tau)\,e^{-i k \tau}}{2\sqrt{\epsilon_1\,k^3} }
 \left[ 1+i k \tau +\frac{\eta}{4} \left( 1 -2 i k (\tau-\tau_1)-e^{2 i k (\tau-\tau_1)}
 \right)\right]
 \label{intfun1}\,,
 \eea
valid for $\tau_1\le\tau\le\tau_2$. This mode function continuously connects, together with its
first derivative,  with the mode
function (and the Bunch-Davies vacuum) for $\tau\le \tau_1$. We can finally connect
the result of eq \eqref{intfun1} with de Sitter mode function at later times $\tau_2\le\tau\le\tau_0$, imposing continuity of the function and its first derivative at
$\tau=\tau_2$.
 The solution corresponds to Ansatz \eqref{ansfu1} with $A_{(n)}=0$, and the scale-dependent
 functions
${\cal C}_1$ and ${\cal C}_2$ are collected in eqs \eqref{defC1} and \eqref{defC2} of the main text.

%%%%%%%% 
\section{Renormalization of UV divergences}
\label{app_reno}
%%%%%%%%

In this Appendix we briefly discuss a method for adsorbing the UV quadratically-divergent
parts \eqref{anloopr1b} of the loop contributions into the available parameters
of the system, at least at large scales for $p\ll1$. The quantities available for this procedure are the overall amplitude ${\cal P}_{0}$ defined in eq \eqref{lsspec}, 
and the factor $\Pi_{0}$ controlling the scale-dependence
of the tree level spectrum \eqref{simPS}. As stated in the main text,
 we can trust our results
only on a large-scale, small-$p$ regime. (See discussion around eq \eqref{pcon1}.) Expanding the total
power spectrum \eqref{dec_ps} up to quadratic order in  $p$, and including the UV one-loop contributions given in eq \eqref{anloopr1b},
we obtain:
\bea
 \label{dec_ps2}
 {\cal P}_{\rm tot}(p)&=&{\cal P}_0\left(1-
  \frac{5\,\Pi_0\,\Lambda_{\rm UV}^2  \,{\cal P}_0\,}{6\,(1+\Pi_0)}
  \right)
% {L}_{\rm loop}^{(a)}
 -\frac{4{\cal P}_0\,\Pi_0}{3} \left(1- \frac{103\, \Lambda_{\rm UV}^2\,{\cal P}_0}{120\,(1+\Pi_0)}
%-\frac{{\cal P}_0}{4} 
%\frac{1}{ (1+\Pi_0)} \left[
 %\frac{\left( \Lambda_{\rm UV}^2 \right)}{10}
 %\right]
\right)\,{p^2}+{\cal O}(p^4)\,.
 \eea
 The parenthesis contain the UV-divergent loop contributions, suppressed
 by a factor ${\cal P}_0$ with respect to the tree-level terms. 
 Higher loop corrections  give contributions to
eq \eqref{dec_ps2} with powers  higher
than two   in ${\cal P}_0$. In the present one-loop instance, we can trust our results only up to quadratic contributions ${\cal P}_0^2$. We can then adsorb the UV-divergent parts of eq 
 \eqref{dec_ps2} into a redefinition of the bare quantities ${\cal P}_{0}$ and $\Pi_{0}$, which are mapped into measurable quantities ${\cal P}_{\rm ms}$ and $\Pi_{\rm ms}$ at large scales: 
% To this order, we 
 %redefine the two bare quantities ${\cal P}_0$ and ${ \Pi}_0$ into  new ones
\bea
\label{renP0}
{\cal P}_0&\to& {\cal P}_{\rm ms}\left(1+\frac{5\,\Lambda_{\rm UV}^2  \,{ \Pi}_{\rm ms}\,}{6\,(1+\Pi_{\rm ms})}\,{\cal P}_{\rm ms}\right)\,,
\\
\label{renPI0}
\Pi_0&\to& \Pi_{\rm ms}
\left(1
+\frac{\Lambda_{\rm UV}^2 \,\left( 103-100 \Pi_{\rm ms}\right)}{120\,(1+\Pi_{\rm ms})}\,{\cal P}_{\rm ms} 
\right)\,.
\eea
By means of this definition, %which includes the one-loop
%divergences into the available parameters, 
 we  express eq \eqref{dec_ps2} as 
\bea
 \label{dec_ps3}
 {\cal P}_{\rm tot}(p)&=&{\cal P}_{\rm ms} -\frac{4{\cal P}_{\rm ms}}{3} \,\Pi_{\rm ms}
 \,{p^2}+{\cal O}(p^4)+{\cal O}({\cal P}_{\rm ms}^2).
 \eea
 Hence quadratically divergent, one-loop effects get adsorbed into bare quantities. The result is expressed 
  in terms of the  the measurable amplitude ${\cal P}_{\rm ms}$ of the spectrum, and on
 the parameter $\Pi_{\rm ms}$ controlling its scale dependence at very large scales (see the discussion around eq \eqref{smke}). 

\smallskip
It is also interesting to provide a feeling on the size of the quadratic loop corrections, by substituting representative values of the measurable parameters in eqs \eqref{renP0} and \eqref{renPI0}. 
At very large scales, we can fix the dimensionless power spectrum  ${\cal P}_{\rm ms}$ 
to the value 
${\cal P}_{\rm ms}\simeq 10^{-9}$, so to match the normalization of CMB spectrum. Moreover, if we wish to get an enhancement at least of order $10^6$
in the size of the primordial spectrum from large towards small scales -- in order for producing primordial black holes -- we select
the ratio 
 $\Pi_{\rm ms}\,=\,10^3$ (see eq \eqref{defPI}). Eqs  \eqref{renP0} and \eqref{renPI0} become
 \bea
\label{renP02}
{\cal P}_0&\simeq& {\cal P}_{\rm ms}\left(1+\frac{5\,\times 10^{-9} }{6}\,\Lambda_{\rm UV}^2\right)\,,
\\
\label{renPI02}
\Pi_0&\simeq& \Pi_{\rm ms}
\left(1-\frac{5\,\times 10^{-9} }{6}\,\Lambda_{\rm UV}^2\right)\,.
\eea
As explained in the main text -- see comments after equation \eqref{strlop1} -- the quantity $\Lambda_{\rm UV}$ is dimensionless,
being the ratio of the cut-off scale versus the pivot scale $k_\star$  of eq \eqref{defpiva} characterizing  the modes leaving the horizon during
the NSR era. 
 Hence bare and measurable quantities are of comparable size, unless the scale of the  cut-off is very large (by a factor at least of order $10^4 - 10^5$) with respect to the pivot scale. 

\end{appendix}

{\small
%\addcontentsline{toc}{section}{References}
%\bibliographystyle{utphys}

%\bibliographystyle{utcaps}
%\bibliographystyle{kp}

%\bibliography{SYMMETRYrefs}
\providecommand{\href}[2]{#2}\begingroup\raggedright\endgroup

}

\end{document}